# Universal DV-CV interaction mechanism for deterministic generation of entangled hybridity

Mihail S. Podoshvedov[1] and Sergey A. Podoshvedov[2]

[1]*Institute of Physics, Kazan Federal University (KFU), 16a Kremlyovskaya St., Kazan, Russia*
*email:mikepodo6@gmail.com*

[2]*Laboratory of Quantum Information Processing and Quantum Computing, Institute of Natural and Exact Sciences, South Ural State University (SUSU), Lenin Av. 76, Chelyabinsk, Russia*
*email: sapodo68@gmail.com*

**Abstract**

We propose a general approach for the implementation of the hybrid entangled states consisting of continuous variable (CV) and discrete variable (DV) states. Peculiarities of DV-CV interaction mechanism on the beam splitter with arbitrary parameters is key for the for the birth of the entangled hybridity provided that some measurement event is registered in auxiliary mode. We show that the CV states forming entangled state can consist exclusively of either even or odd Fock states. Various input CV states including Schrodinger cat state (SCSs) are used at the input of the beam splitter where they mix with delocalized single photon. We show the hybrid entanglement generation is observed at any values of the experimental parameters used. Degree of the hybrid entanglement is evaluated. Conditions for generating the maximally entangled hybridity are established.

**Keywords** DV-CV interaction mechanism, displaced number states, Schrodinger cat states, even/odd CV states

## 1 Introduction

Many physical systems, involving photons, atoms, ions and superconducting platforms, are used for construction of a quantum computer [1] and quantum processors with tens of qubits have been demonstrated [2,3]. In general, despite the wide variety of the physical systems, quantum information processing (QIP) comes in two different depending on the degree of observable used to extract information from quantum state. If we deal with discrete observable (which implies that its eigenvalues are discrete) then we can say about working with discrete variable (DV) state [4,5]. If an observable has continuous set of the eigenvalues, we refer to continuous variable (CV) state [6-8]. Accordingly, the quantum information processing can also be divided into two areas: DV QIP and CV QIP, depending on the manipulated state.

It is completely natural to combine the two approaches in unified hybrid systems by simultaneous use of discrete and continuous variables [9-13]. The advantage of the CV approach to quantum computing lies in that quantum logic gates based on quantum teleportation [6,14] can be implemented deterministically. However, the gates can be realized with restricted fidelity due to finite squeezing which severely limits the CV approach to solving quantum problems from a practical point of view. It is in conflict to the photonic manipulation procedures, for example to qubit teleportation, which is probabilistic since the Bell state measurement of the optical qubits is always probabilistic [15]. The main difficulty of the photonic QIP is in the implementation of controlled-NOT gate. Suppose we have two photonic qubits: controlling $a_0|H\rangle + a_1|V\rangle$ and target $b_0|H\rangle + b_1|V\rangle$ being the superpostions of the single photon in horizontal/vertical basis with corresponding normalized amplitudes. Implementation of the photonic controlled-NOT gate physically means that a polarization of the target qubit is flipped only if a control qubit is horizontally polarized. This is equivalent to superimposing a $\pi$ phase shift between two polarizations conditioned by the existence of another photon in definite polarization. Undoubtedly, this is a nonlinear effect that could be realized in nonlinear optical media. This phenomena can be realized in third-order nonlinear media where the refractive of medium varies with the input light power growing. However currently, no optical material is known in which the nonlinear $\pi$ phase shift could be performed with an acceptable material length taking into account the decoherence effect. Note also that the restriction applies to both DV qubits of different encoding (for example, to superposition of vacuum and single photon $a_0|0\rangle + a_1|1\rangle$) and CV QIP. Note that information of qubits is more robust and can be protected against errors.

Here we consider the possibility of circumventing this limitation due to the peculiarities of the interaction of the CV states with photonic qubits on a beam splitter (BS) without increasing the initial power of the input light states which implies the use of exclusively linear optical elements including photodetectors. We note the nonlinear nature of the interaction which leads to the desired shift by $\pi$ on the target state. We call this type of interaction as a DV-CV nonlinear interaction mechanism on the BS [12,16]. We use this type of interaction to generate the hybrid entangled states consisting of, in general case, even/odd CV states and DV (vacuum, single photon) states. In our case, the output hybrid state is formed from components that are described in Hilbert spaces of various dimensions: finite and infinite, respectively. Interest to the generation of the hybrid states has significantly grown [10, 17-20] since it is believed that the states can be used to resolve the fundamental problems of the QIP by linear optics. Here, we show the possibility of generation of the deterministic entangled hybridity when Schrödinger cat states (SCSs) interacts with delocalized photon on beam splitter. The generation is detected at arbitrary initial parameters (BS parameters, size of the SCSs) and occurs when any measurement outcome is recorded in the auxiliary mode. We show that the CV state containing either even or odd photon states is entangled with the components of the DV qubit (either vacuum or single photon) so that the output entangled state can be described in four dimensional Hilbert space. We also find conditions under which the measure of the entanglement negativity takes on maximally possible values. We develop this approach and show how the DV-CV nonlinear interaction mechanism can be used for deterministic generation of the hybrid entangled states in the case of the CV states other than SCSs. We also consider the interaction of truncated version of the CV state living in a finite Hilbert space with delocalized photon which confirms the possibility of generating the entangled states from Hilbert spaces of different dimensions. The theory being developed is exact and is based on the use of the displaced photon states [21,22] which are, although key concepts for explaining the DV-CV interaction mechanism, but nonetheless they are intermediate for the final results. Interest to the displaced states is only growing [23-28] as an additional classical degree of freedom can be used. In particular, the calculation technique with the displaced states as intermediate is used to search for optimal strategies for generating large-scale SCSs [25] and manipulating these states [26,27].

## 2. DV-CV interaction mechanism
### 2.1 Interaction of SCS and delocalized photon

Consider the interaction of the DV state being the delocalized photon occupying simultaneously modes 2 and 3

$$|\varphi\rangle_{23} = a_0|01\rangle_{23} + a_1|10\rangle_{23}, \tag{1}$$

where amplitudes $a_0$ and $a_1$ satisfy the normalization condition $|a_0|^2 + |a_1|^2 = 1$, with the CV state in mode 1 being even SCS with real amplitude $\beta$. The state is a superposition of two coherent states with different amplitudes in sign $|\pm\beta\rangle$

$$|\beta_+\rangle_1 = N_{ev}^{(0)}(|-\beta\rangle_1 + |\beta\rangle_1), \tag{2}$$

where $N_{ev}^{(0)} = \left(2(1 + exp(-2\beta^2))\right)^{-1/2}$ is the normalization parameter. The beam splitter (BS), which is described by the following unitary matrix

$$BS = \begin{bmatrix} t & -r \\ r & t \end{bmatrix}, \quad (3)$$

where $t$ and $r$ are the real transmittance and reflectance coefficients ($t > 0, r > 0$) respectively, satisfying the normalization condition $t^2 + r^2 = 1$ is used to mix SCS with delocalized photon as shown in Fig. 1(a). Here we used the notation $N_{ev}^{(0)}$ for the normalization factor of the even SCS different from the generally accepted whose meaning will be clear below. Note that we will get similar results if we make use of the odd SCS

$$|\beta_-\rangle_1 = N_{odd}^{(0)}(|-\beta\rangle_1 - |\beta\rangle_1), \quad (4)$$

where $N_{odd}^{(0)} = \left(2(1 - exp(-2\beta^2))\right)^{-1/2}$ is the normalization parameter for odd SCS.

Suppose modes 1 and 2 are mixed on the BS as in Fig. 1(a), then output state can be written as

$$BS_{12}(|\beta_+\rangle_1|\varphi\rangle_{23}) = N_{ev}^{(0)}(BS_{12}(|-\beta\rangle_1|\varphi\rangle_{23}) + BS_{12}(|\beta\rangle_1|\varphi\rangle_{23})), \quad (5)$$

due to linearity of the BS operator. Consider separately each of the terms in Eq. (5). Then we have

$BS_{12}(|-\beta\rangle_1|\varphi\rangle_{23}) = BS_{12}(D_1(-\beta)|0\rangle_1|\varphi\rangle_{23}) = (BS_{12}D_1(-\beta)BS_{12}^+)BS_{12}(|0\rangle_1|\varphi\rangle_{23}) = D_1(-\beta t)D_2(\beta r)BS_{12}(|0\rangle_1|\varphi\rangle_{23}) = D_1(-\beta t)D_2(\beta r)(a_0|00\rangle_{12}|1\rangle_3 + a_1(r|10\rangle_{12} + t|01\rangle_{12})|0\rangle_3) = D_1(-\beta t)(a_0|0\rangle_1|\beta r\rangle_2|1\rangle_3 + a_1(r|1\rangle_1|\beta r\rangle_2 + t|0\rangle_1|1, \beta r\rangle_2)|0\rangle_3) = F(\beta r)D_1(-\beta t)\sum_{n=0}^{\infty}(a_0 c_{0n}(\beta r)|0\rangle_1|1\rangle_3 + a_1(rc_{0n}(\beta r)|1\rangle_1 + tc_{1n}(\beta r)|0\rangle_1)|0\rangle_3)|n\rangle_2,$ (6)

$BS_{12}(|\beta\rangle_1|\varphi\rangle_{23}) = BS_{12}(D_1(\beta)|0\rangle_1|\varphi\rangle_{23}) = (BS_{12}D_1(\beta)BS_{12}^+)BS_{12}(|0\rangle_1|\varphi\rangle_{23}) = D_1(\beta t)D_2(-\beta r)BS_{12}(|0\rangle_1|\varphi\rangle_{23}) = D_1(\beta t)D_2(-\beta r)(a_0|00\rangle_{12}|1\rangle_3 + a_1(r|10\rangle_{12} + t|01\rangle_{12})|0\rangle_3) = D_1(\beta t)(a_0|0\rangle_1|-\beta r\rangle_2|1\rangle_3 + a_1(r|1\rangle_1|-\beta r\rangle_2 + t|0\rangle_1|1, -\beta r\rangle_2)|0\rangle_3) = F(\beta r)D_1(\beta t)\sum_{n=0}^{\infty}(a_0(-1)^n c_{0n}(\beta r)|0\rangle_1|1\rangle_3 + a_1(r(-1)^n c_{0n}(\beta r)|1\rangle_1 + t(-1)^{n-1} c_{1n}(\beta r)|0\rangle_1)|0\rangle_3)|n\rangle_2,$ (7)

where the unitary displacement operator $D(\beta)$ with amplitude $\beta$ is determined by $D(\beta) = exp(\beta a^+ - \beta^* a)$ with $a$ ($a^+$) being the bosonic annihilation (creation) operators and $BS_{12}^+ BS_{12} = I$ with identical operator $I$ and Hermitian conjugate $BS_{12}^+$. Here, we made use of definition of the displaced number states (DNS) with some amplitude $\alpha$

$$|n, \alpha\rangle = D(\alpha)|n\rangle, \quad (8)$$

being intermediate in calculations. Here, we take advantage of decomposing DNS in the basis of the number states [28]

$$|n, \alpha\rangle = F(\alpha) \sum_{n=0}^{\infty} c_{nm}(\alpha) |m\rangle, \quad (9)$$

where $F = exp(-|\alpha|^2/2)$. This decomposition follows from the fact that the infinite set of Fock states $\{|n\rangle, n = 0,1,2, ..., \infty\}$ is complete and therefore any arbitrary pure state can be represented as a superposition in a given basis with amplitudes $c_{nm}(\alpha)$ calculated as $c_{nm}(\alpha) = exp(|\alpha|^2/2)\langle m|n, \alpha\rangle$ so that the normalization condition $exp(-|\alpha|^2)\sum_{m=0}^{\infty}|c_{nm}(\alpha)|^2 = 1$ is performed for any number $n$. In our case, we deal exclusively with intermediate states: coherent state $|\alpha\rangle \equiv |0, \alpha\rangle$ and displaced single photon $D(\alpha)|1\rangle = |1, \alpha\rangle$ whose amplitudes are given by

$$c_{0n}(\alpha) = \frac{\alpha^n}{\sqrt{n!}}, \quad (10)$$

$$c_{1n}(\alpha) = \frac{\alpha^{n-1}}{\sqrt{n!}}(n - |\alpha|^2). \quad (11)$$

As can be seen from the definition of the amplitudes (10, 11), it can be seen that the following relations connecting $c_{0n}(-\alpha)$ ($c_{1n}(-\alpha)$) and $c_{0n}(\alpha)$ ($c_{1n}(\alpha)$) between each other take place

$$c_{0n}(-\alpha) = (-1)^n c_{0n}(\alpha), \quad (12)$$

$$c_{1n}(-\alpha) = (-1)^{n-1} c_{1n}(\alpha). \quad (13)$$

The relationships (12, 13) are the key to the generation of the hybrid entanglement. Finally, the output state in Fig. 1(a) after the BS becomes

$$BS_{12}(|\beta_+\rangle_1|\varphi\rangle_{23}) = N_{ev}^{(0)}F(\beta r)\sum_{n=0}^{\infty}\big(a_0 c_{0n}(\beta r)(|-\beta t\rangle_1+(-1)^n|\beta t\rangle_1)|1\rangle_3 +$$
$$a_1\big(rc_{0n}(\beta r)(|1,-\beta t\rangle_1 + (-1)^n|1,\beta t\rangle_1) + tc_{1n}(\beta r)(|-\beta t\rangle_1+(-1)^{n-1}|\beta t\rangle_1)\big)|0\rangle_3\big)|n\rangle_2. \quad (14)$$

The coherent components of the SCS (2) simultaneously displace the state in one of the modes of the delocalized photon (1) by an amount equal in magnitude but different in sign ($\pm\beta r$) so that all exact information about the displacement of the mode state disappears. The phase contribution of the states $|-\beta r\rangle$ and $|1,-\beta r\rangle$ differs by one due to their different parity, the effect which is akin to the action of the nonlinear effect (DV-CV nonlinear mechanism of interaction). If a measurement outcome with an even number of photons $n = 2m$ is recorded in the second mode, then the following hybrid entangled state is generated

$$|\Delta_{2m}\rangle_{13} = N_{2m}^{(t)}(a_0|\beta_+\rangle_1|1\rangle_3 + a_1 B_{2m}|\Psi_{2m}\rangle_1|0\rangle_3), \quad (15)$$

with the success probability

$$P_{2m} = F^2(\beta r)|c_{02m}(\beta r)|^2 N_{2m}^{(t)-2}\big(N_{ev}^{(0)}(\beta)/N_{ev}^{(0)}(\beta t)\big)^2. \quad (16)$$

If an odd number of photons $n = 2m + 1$ is measured in the second mode, then another hybrid entangled state is generated

$$|\Delta_{2m+1}\rangle_{13} = N_{2m+1}^{(t)}(a_0|\beta_-\rangle_1|1\rangle_3 + a_1 B_{2m+1}|\Psi_{2m+1}\rangle_1|0\rangle_3), \quad (17)$$

with the success probability

$$P_{2m+1} = F^2(\beta r)|c_{02m+1}(\beta r)|^2 N_{2m+1}^{(t)-2}\big(N_{ev}^{(0)}(\beta)/N_{odd}^{(0)}(\beta t)\big)^2. \quad (18)$$

Here, we introduce the following notations. So, the states $|\Psi_{2m}\rangle$ and $|\Psi_{2m+1}\rangle$ are determined by

$$|\Psi_{2m}\rangle = N_{2m}(|\beta_-\rangle + A_{2m}|1,odd\rangle), \quad (19)$$
$$|\Psi_{2m+1}\rangle = N_{2m+1}(|\beta_+\rangle + A_{2m+1}|1,even\rangle), \quad (20)$$

where even/odd SCSs $|\beta_+\rangle$ and $|\beta_-\rangle$ are given above in Eqs. (2, 4) considering that now the amplitude of the coherent states is $\beta t$. We also introduce the superposition of the displaced single photon states (SDSPS) with amplitude $\beta t$

$$|1,even\rangle = N_{ev}^{(1)}(\beta t)(|1,-\beta t\rangle - |1,\beta t\rangle) = -2F(\beta t)N_{ev}^{(1)}(\beta t)\sum_{m=0}^{\infty}c_{12m}(\beta t)|2m\rangle, \quad (21)$$
$$|1,odd\rangle = N_{odd}^{(1)}(\beta t)(|1,-\beta t\rangle + |1,\beta t\rangle) =$$
$$2F(\beta t)N_{odd}^{(1)}(\beta t)\sum_{m=0}^{\infty}c_{12m+1}(\beta t)|2m+1\rangle, \quad (22)$$

where the normalization factors of the SDSPS are given by $N_{ev}^{(1)}(\beta t) = \big(2(1-\exp(-2|\beta t|^2)(1-4|\beta t|^2))\big)^{-1/2}$ and $N_{odd}^{(1)}(\beta t) = \big(2(1+\exp(-2|\beta t|^2)(1-4|\beta t|^2))\big)^{-1/2}$, respectively. We use the number 1 in the designation of the states $|1,even\rangle$ and $|1,odd\rangle$ in order to show that these CV superpositions differ from known ones (2, 4) since they are formed from displaced single photons $|1,\pm\beta t\rangle$. We also use the notations *even/odd* inside the ket vector to show that these superpositions exclusively contain either an even or an odd number of photons as well as the states $|\beta_+\rangle$ and $|\beta_-\rangle$, respectively. Accordingly, we use notation $N_{ev/odd}^{(0,1)}$ for the normalization coefficients with a subscript showing the parity of the Fock states and with superscript that indicates CV states from which this superposition is formed. Accordingly, we can name the state $|1,even\rangle$ (Eq. (21)) even SDSPS, while the state $|1,odd\rangle$ (Eq. (22)) is logical to call odd SDSPS.

The factors $A_{2m}$ and $A_{2m+1}$ are determined by

$$A_{2m} = \frac{r}{t}\frac{c_{02m}(\beta r)}{c_{12m}(\beta r)}\frac{N_{odd}^{(0)}(\beta t)}{N_{odd}^{(1)}(\beta t)}, \quad (23)$$

$$A_{2m+1} = \frac{r}{t}\frac{c_{02m+1}(\beta r)}{c_{12m+1}(\beta r)}\frac{N_{ev}^{(0)}(\beta t)}{N_{ev}^{(1)}(\beta t)}. \tag{24}$$

Note that the states (19) contains exclusively odd Fock states at any value of the parameter $A_{2m}$ (Eq. (23)) therefore, they are logically called odd CV ones. While the states (20) are infinite superpositions of odd Fock states despite the value of the parameter $A_{2m+1}$ (Eq. (24)), therefore, they are called odd CV states. Nevertheless, pairs of the states $|0,even\rangle$ $|1,even\rangle$ and $|0,odd\rangle$ $|1,odd\rangle$ are not orthogonal to each other ($\langle 1,even|0,even\rangle \ne 0, \langle 1,odd|0,odd\rangle \ne 0$), which entails the following normalization factors $N_{2m} = \left(1 + |A_{2m}|^2 - 4N_{odd}^{(0)}(\beta t)N_{odd}^{(1)}(\beta t)exp(-2|\beta t|^2)((\beta t)^* A_{2m} + \beta t A_{2m}^*)\right)^{-1/2}$, $N_{2m+1} = \left(1 + |A_{2m+1}|^2 + 4N_{ev}^{(0)}(\beta t)N_{ev}^{(1)}(\beta t)exp(-2|\beta t|^2)((\beta t)^* A_{2m+1} + \beta t A_{2m+1}^*)\right)^{-1/2}$ for the states $|\Psi_{2m}\rangle$ (Eq. (19)), and $|\Psi_{2m+1}\rangle$ (Eq. (20)), respectively.

The factors $B_{2m}$ and $B_{2m+1}$ defining entangled properties of the conditional hybridity (more precisely this will be discussed in the next chapter) are written as

$$B_{2m} = t\frac{c_{12m}(\beta r)}{c_{02m}(\beta r)}\frac{N_{ev}^{(0)}(\beta t)}{N_{odd}^{(0)}(\beta t)}N_{2m}^{-1}, \tag{25}$$

$$B_{2m+1} = t\frac{c_{12m+1}(\beta r)}{c_{02m+1}(\beta r)}\frac{N_{odd}^{(0)}(\beta t)}{N_{ev}^{(0)}(\beta t)}N_{2m+1}^{-1}, \tag{26}$$

which also determine overall normalization factors $N_{2m}^{(t)}$ and $N_{2m+1}^{(t)}$ of the conditional states (15, 17)

$$N_{2m}^{(t)} = (|a_0|^2 + |a_1|^2|B_{2m}|^2)^{-1/2}, \tag{27}$$

$$N_{2m+1}^{(t)} = (|a_0|^2 + |a_1|^2|B_{2m+1}|^2)^{-1/2}. \tag{28}$$

Using the derived expressions, it is possible directly to show the normalization condition $\sum_{m=0}^{\infty}(P_{2m} + P_{2m+1}) = 1$ is performed. We show three dimensional plots of the probabilities $P_0, P_1, P_2$ and $P_3$ in dependency on $\beta$ and $t$ for the case of $a_0 = a_1 = 1/\sqrt{2}$ in Fig. 2. In the general case, the dependences of $P_0, P_1, P_2$ and $P_3$ are complex. So, we can say that the success probability $P_0$ obviously prevails in the region of small values $\beta$ for the BS with high reflection $t \to 0$. As can be seen from Figure 2, the probability $P_0$ takes the maximum possible values in the case of $\beta = 0$. The value of $\beta = 0$ will hardly be interesting for the hybrid entanglement, it means mixing the vacuum with the state (1). If the value of $\beta$ takes on values close to 0 but nonetheless $\beta \ne 0$, then we can talk about generating the hybrid entanglement with a high success probability greater than 0.9 in the case of $t$ close to zero. The probability $P_0$ decreases both with an increase in the coefficient of transparency $t$ and especially with an increase in the parameter $\beta$. Other success probabilities $P_1, P_2$ and $P_3$ have wavy shape depending on $\beta$. They reach the greatest values already in the case of $\beta \ne 0$. Another observation is related to the behavior of the functions with the parameter $t$ changed. So even probabilities $P_0, P_2$ take the maximum possible values in the case of highly reflective beam splitter $r \to 1, t \to 0$, while maximum values of the probabilities $P_1, P_3$ are observed for highly transmitting ones with large $t$. We also checked the dependence of the probabilities for other values $a_0, a_1$ different from those used in Figure 2. The results of numerical moderation show that the probabilities qualitatively have the same shape as in Fig. 2 with small deformations.

The nonlinear mechanism of interaction between the CV components and the photon qubit allows one to preserve the parity of the state $rc_{0n}(\beta r)(|1,-\beta t\rangle_1 + (-1)^n|1,\beta t\rangle_1) + tc_{1n}(\beta r)(|-\beta t\rangle_1+(-1)^{n-1}|\beta t\rangle_1)$ in the expression (14). Due to this, this state is orthogonal to another CV component $c_{0n}(\beta r)(|-\beta t\rangle_1+(-1)^n|\beta t\rangle_1)$ of the hybrid state for any values $\beta$, $t$ and $n$. Otherwise, in absence of the DV-CV interaction mechanism, these components would not be orthogonal. This mechanism can be extended to other interactions between DV and CV states. It is worth noting that the odd SCS (4) can also be used in the optical scheme in Fig. 1 for conditional generation of entangled hybridity.

## 2.2. Interaction of SCS and two delocalized photons

Due to its versatility, the DV-CV mechanism can be used to generate another type of the hybrid entanglement. This type of the hybrid state involves entanglement between CV states and a delocalized photon simultaneously occupying two modes. To do it let us consider the optical scheme in Fig. 1(b), where now another target state

$$|\varphi'\rangle_{3456} = a_0|0101\rangle_{3456} + a_1|1010\rangle_{3456}, \quad (29)$$

is used. The state is a state of two delocalized photons occupying four modes. Two beam splitters $BS_{13}$ and $BS_{24}$ in Eq. (3) are used in Fig. 1(b) to generate the new type of entangled hybridity. The beam splitter $BS_{13}$ mixes the state (2) with third mode while auxiliary coherent state $|-\beta_1\rangle_2$ with real amplitude $\beta_1 > 0$ interacts with fourth mode of the state (29) on the beam splitter $BS_{24}$. The beam splitter $BS_{24}$ is determined with real transmittance and reflectance coefficients $t_1$ and $r_1$, respectively. To generate the conditional states, two measurements are used in auxiliary modes 3 and 4. Note that the interaction with the coherent state $|-\beta_1\rangle_2$ is optional and is used only to generate the hybrid state with the delocalized photon. The corresponding mathematical calculations are presented in Appendix A. Consider the case of $r_1 \to 0$ and $t_1 \to 1$.

If the measurement outcomes $n = 2m$ and $k$ in modes 3 and 4, respectively, are registered (Fig. 1(b)), then the following conditional state is generated

$$|\Delta_{2mk}\rangle_{123} = N_{2mk}^{(t)}(a_0|\beta_+\rangle_1|01\rangle_{23} + a_1 B_{2mk}|\Psi_{2mk}\rangle_1|10\rangle_{23}), \quad (30)$$

If the measurement outcomes $n = 2m + 1$ and $k$ in modes 3 and 4 are fixed (Fig. 1(b)), then the following entangled hybrid state appears

$$|\Delta_{2m+1k}\rangle_{123} = N_{2m+1k}^{(t)}(a_0|\beta_-\rangle_1|01\rangle_{23} + a_1 B_{2m+1k}|\Psi_{2m+1k}\rangle_1|10\rangle_{23}). \quad (31)$$

All other parameters are given in Appendix A. Here we have replaced the designation of modes 5 and 6 for the delocalized photon by 1 and 2 ($5 \to 2, 6 \to 3$). To obtain the states (30, 31), we believe that $|\Psi_k\rangle_2 \approx |-\beta_1 r_1\rangle_2$ that is correct with high fidelity in the case of $r_1 \to 0$ and $t_1 \to 1$. The success probabilities to generate the states can be written as

$$P_{2mk} = F^2(\beta r)F^2(\beta_1 r_1)|c_{02m}(\beta r)|^2\left(N_{2mk}^{(t)}N_k\right)^{-2}\left(N_{ev}^{(0)}(\beta)/N_{ev}^{(0)}(\beta t)\right)^2, \quad (32)$$

$$P_{2m+1k} = F^2(\beta r)F^2(\beta_1 r_1)|c_{02m+1}(\beta r)|^2\left(N_{2m+1k}^{(t)}N_k\right)^{-2}\left(N_{ev}^{(0)}(\beta)/N_{odd}^{(0)}(\beta t)\right)^2. \quad (33)$$

It is possible directly to check the probabilities (32, 33) are normalized $\sum_{m=0}^{\infty}\sum_{k=0}^{\infty}(P_{2mk} + P_{2m+1k}) = 1$.

The dependencies of the probabilities $P_{00}$ (measurement of vacuum in both auxiliary third and fourth modes), $P_{01}$ (measurement of vacuum in third and single photon in fourth modes), $P_{10}$ (measurement of single photon in third and vacuum in fourth modes) and $P_{11}$ (measurement of the single photons in both third and fourth auxiliary modes) on the parameters of $\beta$ and $t$ are shown in Figure 3 in the case of $a_0 = a_1 = 1/\sqrt{2}, \beta_1 = 1, t_1 = 0.95$. These dependencies have some similarities with those presented in Fig. 2. So, fairly high success probabilities of $P_{00}$ and $P_{01}$ (close to 0.5) are observed in the region of $\beta \approx 0$. Sufficiently high values of the probability of $P_{01}$ are observed for any values of the transparency coefficient $t$, in contrast to $P_{00}$ which falls with increasing $t$. The contribution of the probabilities $P_{10}$ and $P_{11}$ is absolutely insignificant in the case of small size $|\beta| \leq 1$ for any parameter values of $t$. If the parameter $|\beta|$ begins to increase, then the contribution of the success probabilities $P_{10}$ and $P_{11}$ increases, especially with $t$ growing. We only note that the range of changes for $P_{10}$ and $P_{11}$ (in range of $\leq 0.12$) is less than the range of variation of $P_{00}$ and $P_{01}$ which suggests that the contribution of events $P_{00}$ and $P_{01}$ prevails over all other ones.

Note that one can also consider the generation of an entangled state of light in Figure 1(b) without using a second beam splitter $BS_{24}$. The nonlinear nature of the DV-CV interaction also allows the generation of an entangled hybridity of the CV state with a rather exotic DV state $|101\rangle + |010\rangle$. In this case, it will be possible to use the optical scheme in Fig. 1(a) and the results presented in the previous section, except for the fact that the DV state will be different from superposition of vacuum and single photon.

We also note that if we use the conversion of the which-path encoding $\{|01\rangle, |10\rangle\}$, of the single photon in a polarization basis with horizontal $|H\rangle$ and vertical $|V\rangle$ polarizations which can be done by linear optics, then the states (30, 31) are transformed to $|\Delta_{2mk}\rangle_{12} = N_{2mk}^{(t)}(a_0|\beta_+\rangle_1|H\rangle_2 + a_1 B_{2mk}|\Psi_{2mk}\rangle_1|V\rangle_2)$ and $|\Delta_{2m+1k}\rangle_{12} = N_{2m+1k}^{(t)}(a_0|\beta_-\rangle_1|H\rangle_2 + a_1 B_{2m+1k}|\Psi_{2m+1k}\rangle_1|V\rangle_2)$, respectively.

## 2.3. Interaction of truncated SCSs with photonic qubit

Generation of the SCSs, especially of a large size, in free-propagating light is a complex problem of technically unsolved so far. As a rule, the research efforts are aimed at generating a finite superposition which with some fidelity is truncated version of SCSs. However, these truncated versions of the SCSs can be useful for practical applications. Consider the manifestation of the studied DV-CV interaction mechanism on the example of interaction of the truncated version of the SCSs with, for example, the state (1). In the case, in essence, we are talking about the interaction of two DV states between each other. The difference between the DV states is that they can be defined in Hilbert spaces of various dimensions. Consider a truncated version of even SCS in (2) which is superposition of even number states living in $n+1$ Hilbert space

$$|\Sigma_{in}^{(n)}\rangle = N_n \sum_{k=0}^{n} \frac{\beta^{2k}}{\sqrt{(2k)!}} |2k\rangle, \tag{34}$$

where the parameter $N_n$ is the normalization factor and $n$ stands for the number of terms in the truncated version of SCS. It is interesting to trace whether it is possible to observe the manifestation of the DV-CV interaction mechanism on the truncated version of the SCS in Fig. 1(a).

So, if we consider truncated version $|\Sigma_{in}^{(2)}\rangle = N_2(|0\rangle + (\beta^2/\sqrt{2!})|2\rangle)$, then the following conditional states are generated

$$|\Omega_0^{(2)}\rangle_{13} = N_0^{(2)}\left(a_0\left(|0\rangle_1 + \frac{\beta^2 t^2}{\sqrt{2!}}|2\rangle_1\right)|1\rangle_3 + a_1\sqrt{1-t^2}\left(|1\rangle_1 + \sqrt{\frac{3}{2}}\beta^2 t^2|3\rangle_1\right)|0\rangle_3\right), \tag{35}$$

provided that vacuum state is fixed in auxiliary second mode and

$$|\Omega_1^{(2)}\rangle_{13} = N_1^{(2)}\left(-a_0 t\sqrt{1-t^2}\beta^2|1\rangle_1|1\rangle_3 + a_1 t\left(|0\rangle_1 + (t^2-2r^2)\frac{\beta^2}{\sqrt{2!}}|2\rangle_1\right)|0\rangle_3\right), \tag{36}$$

provided that single photon is registered in the second mode. Here, $N_0^{(2)}$ and $N_1^{(2)}$ are the corresponding normalization factors $N_0^{(2)} = \left(|a_0|^2(1+\beta^4 t^4/2!) + |a_1|^2(1-t^2)(1+3\beta^4 t^4/2)\right)^{-1/2}$ and $N_1^{(2)} = \left(|a_0|^2\beta^4 t^2(1-t^2) + |a_1|^2 t^2(1+(t^2-2r^2)^2\beta^4/2)\right)^{-1/2}$. The subscript is responsible for the number of photons measured in second auxiliary mode, while the superscript indicates that the conditional hybrid state was generated from truncated version of SCS with corresponding number of terms.

Consider another version of the truncated SCS containing first three terms $|\Sigma_{in}^{(3)}\rangle = N_3(|0\rangle + (\beta^2/\sqrt{2!})|2\rangle + (\beta^4/\sqrt{4!})|4\rangle)$, where $N_3$ is the normalization factor. Then, the following conditional hybrid states are produced

$$|\Omega_0^{(3)}\rangle_{13} = N_0^{(3)}\left(\begin{array}{l} a_0\left(|0\rangle_1 + \frac{\beta^2 t^2}{\sqrt{2!}}|2\rangle_1 + \frac{\beta^4 t^4}{\sqrt{4!}}|4\rangle_1\right)|1\rangle_3 + \\ a_1\sqrt{1-t^2}\left(|1\rangle_1 + \sqrt{\frac{3}{2}}\beta^2 t^2|3\rangle_1 + \sqrt{\frac{5}{4!}}\beta^4 t^4|5\rangle_1\right)|0\rangle_3 \end{array}\right), \tag{37}$$

$$|\Omega_1^{(3)}\rangle_{13} = N_1^{(3)} \begin{pmatrix} -a_0 t\sqrt{1-t^2}\beta^2 \left(|1\rangle_1 + \frac{\beta^2 t^2}{\sqrt{3!}}|3\rangle_1\right)|1\rangle_3 + \\ a_1 t\left(|0\rangle_1 + (t^2-2r^2)\frac{\beta^2}{\sqrt{2!}}|2\rangle_1 + (t^2-4r^2)\frac{\beta^4 t^2}{\sqrt{4!}}|4\rangle_1\right)|0\rangle_3 \end{pmatrix}, \quad (38)$$

provided that the corresponding measurement outcomes are registered in second auxiliary mode. The quantities $N_0^{(3)}$ and $N_1^{(3)}$ are the normalizations factors. Note that the states (35-38) in the first mode contain either only even or only an odd number of photons but not both at once what is consistent to above results. The states in the first mode are orthogonal to each other. This proves that the nonlinear DV-CV interaction mechanism is also applicable to the truncated versions of SCS interacting with DV states and enables to produce entangled states between two DV states. Consideration of the conditional states can be continued provided that truncated versions of SCS (34) with larger number of superposition terms are used.

Now consider how the generated states can be close to the hybrid states resulting from the application of genuine SCS (2). As a rule, the parameter fidelity is used for this purpose, which allows us to evaluate the closeness of the states to each other. For two pure states, the fidelity is defined as $F_0^{(n)} = \left|\langle \Omega_0^{(n)}|\Delta_0\rangle\right|^2$ and $F_1^{(n)} = \left|\langle \Omega_1^{(n)}|\Delta_1\rangle\right|^2$, when vacuum and single photon, respectively, are registered in auxiliary second mode. The mathematical expressions used for the numerical calculation of the fidelities are presented in Appendix B. In Fig. 4 we show the dependence of the fidelities $F_0^{(2)}$ (Fig. 4(a)), $F_0^{(3)}$ (Fig. 4(b)), $F_1^{(2)}$ (Fig. 4(c)) and $F_1^{(3)}$ (Fig. 4(d)) on the parameter $\beta$ for various values of the transmittance $t$ of the beam splitter. Since the CV components of the states (15, 17) have a size $\beta t$, then SCS (Eq. (2)) of large size $\beta$ is required to generate the hybrid entanglement for small values of the parameter $t$, which is expressed in the fact that the fidelities $F_0^{(2)}$, $F_0^{(3)}$, $F_1^{(2)}$ and $F_1^{(3)}$ can take on sufficiently large values with $\beta$ growing in the case of small values of $t$. As can be seen from the plots, the fidelity between states in models with truncated and genuine SCS in Fig. 1(a) takes almost ideal value 1 in a wide range of experimental parameters, which confirms the possibility of using the states (34) in producing entangled hybrid states.

## 3. Entanglement of the generated hybridity

The conditional states (15, 17, 30, 31) are entangled. To evaluate the measure of state entanglement, we will use negativity [29] which can be easy computed in a four-dimensional Hilbert space. This quantity can be derived from the positive partial transpose (PPT) criterion for separability [30] and has all required properties for the entanglement measure. The negativity $\mathcal{N}$ of a bipartite composed system $AB$ characterized by a density matrix $\varrho$ is defined by $\mathcal{N} = (\|\varrho^{T_A}\| - 1)$, where $\varrho^{T_A}$ is the partial transpose of $\varrho$ with respect to the subsystem $A$ and $\|\varrho^{T_A}\| = Tr|\varrho^{T_A}| = Tr\sqrt{(\varrho^{T_A})^+\varrho^{T_A}}$ is the trace norm of the sum of the singular values of the operator $\varrho^{T_A}$, where $(\varrho^{T_A})^+$ means Hermitian conjugate operator of original $\varrho^{T_A}$. The negativity value ranges from $\mathcal{N} = 0$ (separable state) up to $\mathcal{N}_{max} = 1$ (maximally entangled state).

First, evaluate the measure of entanglement of the states (15, 17). The states $|\beta_+\rangle$ and $|\Psi_{2m}\rangle$ being CV components of the state (15) are orthogonal to each other $\langle\beta_+|\Psi_{2m}\rangle = 0$ regardless of the value of the parameter $A_{2m}$ (Eq. (23)). It follows from the fact the state $|\beta_+\rangle$ contains only even Fock states, while the state $|\Psi_{2m}\rangle$ is infinite superposition of odd number states. Therefore, the conditional state (15) can be defined in a four-dimensional Hilbert space with the basic states $\{|\beta_+\rangle_1|1\rangle_2, |\beta_+\rangle_1|0\rangle_2, |\Psi_{2m}\rangle_1|1\rangle_2, |\Psi_{2m}\rangle_1|0\rangle_2\}$. The same approach applies to the analysis of the entangled state (17) whose CV components are also orthogonal $\langle\beta_-|\Psi_{2m+1}\rangle = 0$ for any value of the parameter $A_{2m+1}$ (Eq. (24)). The state $|\beta_-\rangle$ involves only odd Fock states and $|\Psi_{2m+1}\rangle$ is only composed of even number state. The state (17) can be described in a four-dimensional Hilbert space with the base states $\{|\beta_-\rangle_1|1\rangle_2, |\beta_-\rangle_1|0\rangle_2, |\Psi_{2m+1}\rangle_1|1\rangle_2, |\Psi_{2m+1}\rangle_1|0\rangle_2\}$. Calculations give the following results

$$\mathcal{N}_{2m} = \frac{2|a_0||a_1||B_{2m}|}{|a_0|^2+|a_1|^2|B_{2m}|^2}, \tag{39}$$

$$\mathcal{N}_{2m+1} = \frac{2|a_0||a_1||B_{2m+1}|}{|a_0|^2+|a_1|^2|B_{2m+1}|^2}. \tag{40}$$

The same approach applies to states (30, 31) in calculating their negativities. Indeed, these states can also be defined in the four-dimensional Hilbert space regardless of the parameter values $A_{2mk}(A_{2m+1k})$ (Eqs. (23, 24)) and $B_{2mk}(B_{2m+1k})$. Pairs of the CV states $(|\beta_+\rangle, |\Psi_{2mk}\rangle)$ and $(|\beta_-\rangle, |\Psi_{2m+1k}\rangle)$ are orthogonal to each other as well as DV states $(|01\rangle, |10\rangle)$. This allows us to define a four-dimensional Hilbert space as $\{|even\rangle_1|01\rangle_2, |even\rangle_1|10\rangle_2, |odd\rangle_1|01\rangle_2, |odd\rangle_1|10\rangle_2\}$, where $|even/odd\rangle$ states mean all CV states consisting of either even or odd Fock states. Following the same technique, we have the negativities

$$\mathcal{N}_{2mk} = \frac{2|a_0||a_1||B_{2mk}|}{|a_0|^2+|a_1|^2|B_{2mk}|^2}, \tag{41}$$

$$\mathcal{N}_{2m+1k} = \frac{2|a_0||a_1||B_{2m+1k}|}{|a_0|^2+|a_1|^2|B_{2m+1k}|^2}. \tag{42}$$

Corresponding three-dimensional dependencies of the negativities $\mathcal{N}_0$, $\mathcal{N}_1$, $\mathcal{N}_{00}$ and $\mathcal{N}_{10}$ on the parameters $\beta$ and $t$ are shown in Fig. 5 and 6, respectively. We present these dependencies as the most typical observed by us. A high quantity of the hybrid entanglement is observed for the states $|\Delta_0\rangle$ and $|\Delta_{00}\rangle$ in almost entire range of the parameters $\beta$ and $t$. For other states in Figs. 5 and 6, there are areas, where the negativity takes much smaller values down to zero, especially near $\beta \approx 0$. In general, the use of additional experimental parameters $\beta_1$ and $t_1$ in Fig. 1(b) compared to the case in Fig. 1(a) may expand the possibilities for generating the entangled hybridity. A deep ditch along the straight line $\beta = 0$ for the negativities $\mathcal{N}_k = 0$ and $\mathcal{N}_{kl} = 0$ with $k > 0$ is due to the fact that in this case no entanglement is generated. The lack of the entanglement in the case can be verified by direct calculation. The negatives $\mathcal{N}_0$ and $\mathcal{N}_{00}$ take nonzero values even in the case of $\beta = 0$ that indicates about the generation of the entangled (but not hybrid) states. Both negativities $\mathcal{N}_{2m}, \mathcal{N}_{2m+1}$ and $\mathcal{N}_{2mk}, \mathcal{N}_{2m+1k}$ depend on the parameters $B_{2m}, B_{2m+1}$ and $B_{2mk}, B_{2m+1k}$, respectively, the parameters depends on the decomposition amplitude of the displaced single photon $c_{1n}(\beta r)$ (Eq. (11)). The amplitude becomes equal to zero if either $\beta r = \sqrt{2m}$ or $\beta r = \sqrt{2m+1}$ for the case of $\beta \neq 0$ and therefore, one might think that negativity also takes on zero values for given values. Nevertheless, the parameters $B_{2m}, B_{2m+1}$ and $B_{2mk}, B_{2m+1k}$ also contain additional factors $N_{2m}^{-1}, N_{2m+1}^{-1}$ and $N_{2mk}^{-1}, N_{2m+1k}^{-1}$ that grade the contribution of $c_{1n}(\beta r)$. Finally, the negativities (39-42) never take zero values except of the case of $\beta = 0$ not being an example of the generation of entangled hybridity. That is why the generation of the hybrid entanglement in Figs. 1(a, b) is deterministic.

Conversely, the maximum negativity $\mathcal{N}_{max} = 1$ of the generated states can be determined from the condition $B_{2m}(B_{2m+1}) = B_{2mk}(B_{2m+1k}) = \sqrt{|a_0|/|a_1|}$. It follows from it the set of initial parameters $(|a_0|, |a_0|, \beta, t)$ for the case in Fig.1(a) and $(|a_0|, |a_0|, \beta, t, \beta_1, t_1)$ for the case in Fig.1(b) defines measure of the hybrid entanglement. The maximum negativity for balanced $(|a_0| = |a_0| = 1/\sqrt{2})$ superposition (1) is observed under such settings which provide $B_{2m}(B_{2m+1}) = B_{2mk}(B_{2m+1k}) = 1$. If the experimental parameters are chosen in such a way that $B_{2m}(B_{2m+1}) = B_{2mk}(B_{2m+1k}) \neq 1$, then unbalanced $(|a_0| \neq |a_0|)$ DV superposition (1) should be used in Figs. 1(a) and 1(b). In Table 1 we present partial numerical values the experimental settings that can be used to generate maximum hybridity in optical schemes in Fig. 1(a). Note that the settings that guarantee maximum entanglement $\mathcal{N}_{max} = 1$ are huge. It is interesting to note that maximum hybrid entanglement is generated both for insignificant SCS amplitude values ($\beta = 0.1$) and those values which are not yet available with perfect fidelity, for example, $\beta = 1.88492$ in the case of $n = 0$. We could even say that, from a practical point of view, generating the hybrid entanglement with a small value of $\beta$ (for example, $\beta = 0.1$) may even be of more interest in practice. The success probability of generation of such maximally entangled hybridity is high enough ($P_0 = 0.969829$ for $\beta = 0.1$) and SCS of such small size can be realized into practice. The generation of maximum hybrid entanglement is also observed for the case in the optical scheme in Figure 1(b) both in the case of use of balanced and unbalanced superposition (29) (numerical data are not presented). In addition, in this case, two additional parameters $\beta_1$ and $t_1$ can be used to manipulate the output hybrid entanglement.

| $n$ | $|a_0|$ | $|a_1|$ | $\beta$ | $t$ | $P_0, P_1$ |
|---|---|---|---|---|---|
| 0 | $1/\sqrt{2}$ | $1/\sqrt{2}$ | 1.88492 | 0.3 | 0.0394327 |
|   |              |              | 1.56391 | 0.5 | 0.159716 |
|   |              |              | 1.70713 | 0.8 | 0.350236 |
|   | $\sqrt{0.4796}$ | $\sqrt{0.5204}$ | 0.1 | 0.2 | 0.969829 |
|   | $\sqrt{0.36}$ | $\sqrt{0.64}$ | 0.1 | 0.5 | 0.833727 |
|   | $\sqrt{0.11486}$ | $\sqrt{0.88514}$ | 0.2 | 0.8 | 0.427518 |
|   | $\sqrt{0.4797}$ | $\sqrt{0.5203}$ | 0.6 | 0.2 | 0.693138 |
| 1 | $1/\sqrt{2}$ | $1/\sqrt{2}$ | 1.26429 | 0.8 | 0.270754 |
|   | $1/\sqrt{2}$ | $1/\sqrt{2}$ | 1.47621 | 0.9 | 0.262298 |
|   | $\sqrt{0.97767}$ | $\sqrt{0.02233}$ | 0.8 | 0.8 | 0.09011 |
|   | $\sqrt{0.96389}$ | $\sqrt{0.03611}$ | 1 | 0.9 | 0.127707 |
|   | $\sqrt{0.85578}$ | $\sqrt{0.14422}$ | 1.2 | 0.9 | 0.214779 |
|   | $\sqrt{0.44517}$ | $\sqrt{0.55483}$ | 1.3 | 0.8 | 0.266999 |

**Table 1.** The initial settings $(|a_0|, |a_0|, \beta, t)$ guarantee the generation of the maximum hybrid entanglement $\mathcal{N}_{max} = 1$ in the optical scheme in Figure 1(a) with corresponding success probabilities $P_0$, $P_1$ in the case of registration of vacuum and single photon in auxiliary second mode.

Note that this consideration is applicable to the calculation of the negativity for the entangled states (35-38). The states can also be defined in four-dimensional Hilbert space $\{|even\rangle_1|0\rangle_2, |even\rangle_1|1\rangle_2, |odd\rangle_1|0\rangle_2, |odd\rangle_1|1\rangle_2\}$ regardless of their amplitudes. In this case, the notation $|even/odd\rangle$ means all DV states that are formed from either even or odd Fock states. Then, negativity is expressed by the same expressions (39-42) with the entangling parameters $B_0^{(2)} = \sqrt{1-t^2}\sqrt{(1+3\beta^4 t^4/2)/(1+\beta^4 t^4/2)}$ for the state (35), $B_0^{(3)} = \sqrt{1-t^2}\sqrt{(1+3\beta^4 t^4/2 + 5\beta^8 t^8/4!)/(1+\beta^4 t^4/2 + \beta^8 t^8/4!)}$ for the state (37), $B_1^{(2)} = \beta^{-2}\sqrt{(1+\beta^4(t^2-2r^2)^2/2)/(1-t^2)}$ for the state (36) and $B_1^{(3)} = \beta^{-2}\sqrt{(1+\beta^4(t^2-2r^2)^2/2 + \beta^8 t^2(t^2-4r^2)^2/4!)/(1-t^2)(1+\beta^4 t^4/3!)}$ for the state (38). Using these expressions, one can find the values of the experimental parameters $(\beta, t)$ at which maximally entangled DV-DV states are generated.

## 4. DV-CV interaction mechanism on example of the states $|\Psi_{2m}\rangle$ and $|\Psi_{2m+1}\rangle$

In the previous section, we showed the possibility of generating the hybrid entanglement in the case of use of even SCS (2). It is interesting to check whether the generation of the entangled hybridity is possible in the case of using other CV states, for example, in the case of use of the states $|\Psi_{2m}\rangle$ (Eq. (19)). Consider the case of interaction of the state $|\Psi_{2m}\rangle$ (Eq. (19)) with delocalized photon (1). For this purpose, Fig. 1 (a) can be used in which the SCS is replaced by the CV state $|\Psi_{2m}\rangle$ in first input mode. For simplicity, we consider the state $|\Psi_{2m}\rangle$ with an amplitude $\beta$ (instead of $\beta t$ as in Eq. (19)) and also use the notation $A_{2m}$ for amplitude of $|1, odd\rangle$, nevertheless, implying its arbitrary value.

The DV-CV nonlinear interaction mechanism turns out to work in this case (Appendix C). Using analytical expressions (C1-C6), it is finally possible to show that the optical scheme in Fig. 1(a) enables to generate the following hybrid entangled states

$$|\Phi_{2l}\rangle_{13} = N_{2l}^{(t)}\left(a_0|\Psi_{odd}^{(2l)}\rangle_1 |1\rangle_3 + a_1 B_{2l}|\Psi_{ev}^{(2l)}\rangle_1 |0\rangle_3\right), \quad (43)$$

provided that even number of photons $2l$ is measured in auxiliary second mode and

$$|\Phi_{2l+1}\rangle_{13} = N_{2l+1}^{(t)}\left(a_0|\Psi_{ev}^{(2l+1)}\rangle_1 |1\rangle_3 + a_1 B_{2l+1}|\Psi_{odd}^{(2l+1)}\rangle_1 |0\rangle_3\right), \quad (44)$$

provided that odd number of photons $2l + 1$ is measured in auxiliary second mode. Here, the CV states forming the entangled hybrid states are the following

$$|\Psi_{odd}^{(2l)}\rangle = N_{odd}^{(2l)}(|\beta_-\rangle + C_{2l}|1, odd\rangle), \quad (45)$$

$$|\Psi_{ev}^{(2l)}\rangle = N_{ev}^{(2l)}(|\beta_+\rangle + D_{2l}|1, even\rangle + F_{2l}|2, even\rangle), \quad (46)$$

$$|\Psi_{ev}^{(2l+1)}\rangle = N_{ev}^{(2l+1)}(|\beta_+\rangle + C_{2l+1}|1, even\rangle), \quad (47)$$

$$|\Psi_{odd}^{(2l+1)}\rangle = N_{odd}^{(2l+1)}(|\beta_-\rangle + D_{2l+1}|1, odd\rangle + F_{2l+1}|2, odd\rangle). \quad (48)$$

Again, as in the cases considered above, the CV states consist of either exclusively even $\left(|\Psi_{ev}^{(2l)}\rangle, |\Psi_{ev}^{(2l+1)}\rangle\right)$ or odd $\left(|\Psi_{odd}^{(2l)}\rangle, |\Psi_{odd}^{(2l+1)}\rangle\right)$ Fock states regardless of the parameter values $C_{2l}/C_{2l+1}$, $D_{2l}/D_{2l+1}$ and $F_{2l}/F_{2l+1}$. Note also that states (45-48) include additional terms. So the states (45, 47) additionally include a superposition of displaced single photons (either $|1, odd\rangle$ or $|1, even\rangle$), in contrast to the case (15, 17) with SCS as an input in Fig. 1(a). Also other CV states (46, 48) include an additional state superposition of displaced two photon states (C7) unlike the CV state with input SCS in Fig. 1(a). Analytical expressions for coefficients $C_{2l}$ (C5), $D_{2l}$ (C6) and $F_{2l}$ (C7) are presented in Appendix C. The coefficients $C_{2l+1}$, $D_{2l+1}$ and $F_{2l+1}$ for the states (47, 48) can be determined by analogy.

The entangled hybrid states (43, 44) can also be considered in four-dimensional Hilbert space despite CV nature of the states (45-48) in first mode. The states (45-48) can be defined in a two-dimensional Hilbert space with basic states, either even or odd CV state. We note this representation of CV states in a two-dimensional Hilbert space is possible despite the fact that each of these states is determined by a continuous variable. DV state is also defined in two-dimensional Hilbert space with base states either $|0\rangle$ or $|1\rangle$. Then, the base of four-dimensional Hilbert space becomes $\{|even\rangle_1|0\rangle_2, |even\rangle_1|1\rangle_2, |odd\rangle_1|0\rangle_2, |odd\rangle_1|1\rangle_2\}$ and negativity can be calculated with help of the expressions (39-42), where now entanglement parameter is given by

$$B_{2l} = \frac{N_{odd}^{(2l)}N_{odd}^{(0)}(\beta t)\left(tc_{12l}(\beta r)N_{odd}^{(0)}(\beta) - \sqrt{2}trA_{2m}c_{22l}(\beta r)N_{odd}^{(1)}(\beta)\right)}{N_{ev}^{(2l)}N_{ev}^{(0)}(\beta t)R_{2l}}, \quad (49)$$

$$B_{2l+1} = \frac{N_{ev}^{(2l+1)}N_{ev}^{(0)}(\beta t)\left(tc_{12l}(\beta r)N_{odd}^{(0)}(\beta) - \sqrt{2}trA_{2m}c_{22l}(\beta r)N_{odd}^{(1)}(\beta)\right)}{N_{odd}^{(2l+1)}N_{odd}^{(0)}(\beta t)R_{2l+1}}, \quad (50)$$

where we introduce the following normalization factors: $N_{odd}^{(2l)}$ is for the state (45), $N_{ev}^{(2l)}$ is for the state (46), $N_{ev}^{(2l+1)}$ is for the state (47) and $N_{odd}^{(2l+1)}$ is for the state (48). The analytical expressions for the normalization factors are rather complex (therefore, they are not presented here) since the states $|\beta_+\rangle$ ($|\beta_-\rangle$), $|1, even\rangle$ ($|1, odd\rangle$), $|2, even\rangle$ ($|2, odd\rangle$) are not orthogonal to each other. As a result, the normalization factors will contain nonzero cross factors like $\langle\beta_-|1, odd\rangle$. And vice versa, the CV components of the states (43, 44) are orthogonal to each other, therefore the normalization factors $N_{2l}^{(t)}$ and $N_{2l+1}^{(t)}$ can be represented in the form

$$N_{2l}^{(t)} = (|a_0|^2 + |a_1|^2|B_{2l}|^2)^{-1/2}, \tag{51}$$

$$N_{2l+1}^{(t)} = (|a_0|^2 + |a_1|^2|B_{2l+1}|^2)^{-1/2}. \tag{52}$$

Using the parameters, we can finally write an expression for the probability of generating the corresponding hybrid state

$$P_{2l} = F^2(\beta r)\frac{N_{2m}^2 R_{2l}^2}{N_{odd}^{(0)2}(\beta t)N_{odd}^{(2l)2}N_{2l}^{(t)2}}, \tag{53}$$

$$P_{2l+1} = F^2(\beta r)\frac{N_{2m}^2 R_{2l+1}^2}{N_{ev}^{(0)2}(\beta t)N_{ev}^{(2l+1)2}N_{2l+1}^{(t)2}}. \tag{54}$$

Similar transformations can also be carried out with the initial state $|\Psi_{2m+1}\rangle$ (Eq. (20)) in Fig. 1(a) with rather tedious calculations and the result also will be the generation of an entangled hybrid state with either even or odd CV states in first mode. Thus, we have shown that the CV-DV nonlinear mechanism can be implemented for other CV states being input in Fig. 1(a). The interaction of the CV state with vacuum on the BS preserves the parity of the input CV state, while the interaction of the same state with a single photon changes the parity of the output SM state. This allows one to deterministically generate the hybrid entangled states (43, 44) that can be described in four-dimensional Hilbert space.

## 5. Conclusion

We examined the problem of the interaction of CV states with DV ones on the beam splitter in the most general case for arbitrary values of the experimental parameters. The number of the optical elements in optical schemes in Fig. 1(a, b) is irreducible and number of resources consumed is minimal. We have shown the inevitability of generating the hybrid entangled state in this simple optical scheme due to DV-CV nonlinear interaction mechanism. In particular, we examined the interaction of the even/odd SCSs (2, 4) with the delocalized single photon (1). This type of the DV-CV interaction can be carried out in a more complex form with two delocalized photons (29) and one auxiliary coherent state as shown in Fig. 1(b). We also showed that DV-CV interaction can be realized simply by use of input either even or odd CV states (19, 20). The final result of this type of interaction is the generation of the hybrid entanglement (43, 44) provided that some measurement event is registered in auxiliary mode. Moreover, either even CV (containing exclusively even Fock states) or odd CV (containing exclusively odd Fock states) states become entangled with the DV state. In other words, the CV states are orthogonal to each other regardless of the choice of the experimental parameters, for example, size $\beta$ of the SCSs, parameters of the BS.

The reason for this effect lies in the behavior of the matrix elements (10, 11) being expansion amplitudes in the Fock basis when the displacement amplitude changes to the opposite value (12, 13) depending on the parity of the displaced state. This mechanism enables to maintain the parity of the CV state interacting with DV one either vacuum or single photon. We showed that the deterministic nature of the birth of the hybrid entanglement, that is, an entangled state is generated for any fixed measured outcome. The measure of entanglement can vary over a wide range, but there are also innumerable number of the values of the experimental parameters that provide maximum entanglement of the output hybrid state. We also tested the possibility of implementing the hybrid entanglement if a truncated version of the SCSs (as a more realistic version of the SCSs used in practice) is used at the input. We have confirmed the possibility of generating entangled hybrid states (35-38) with sufficiently high fidelity at certain amplitude values $\beta$ of the SCSs. In this case, we can talk about the generation of the DV-DV entanglement of the states of various dimensions. We have shown that the emerging entangled state can also be described in four-dimensional Hilbert space. Therefore, it is possible to talk about the DV-DV nonlinear interaction mechanism responsible for the birth of the hybrid entanglement.

## Appendix A. Analysis of the optical scheme in Fig. 1(b)

We follow the same mathematical technique that was used to analyze the optical scheme in Figure 1(b). So, we have

$$|\Delta'_{2mk}\rangle_{156} = N^{(t)}_{2mk}(a_0|\beta_+\rangle_1|\Psi_k\rangle_2|01\rangle_{56} + a_1 B_{2mk}|\Psi_{2mk}\rangle_1|-\beta_1 t_1\rangle_2|10\rangle_{56}), \quad (A1)$$

if even number of photons $n = 2m$ and $k$ photons are fixed in third and fourth modes and

$$|\Delta'_{2m+1k}\rangle_{156} = N^{(t)}_{2m+1k}(a_0|\beta_-\rangle_1|\Psi_k\rangle_2|01\rangle_{56} + a_1 B_{2m+1k}|\Psi_{2m+1k}\rangle_1|-\beta_1 t_1\rangle_1|10\rangle_{56}), \quad (A2)$$

if the detectors register odd number of photons $n = 2m + 1$ and $k$ photons in the modes. The states we introduce are the following

$$|\Psi_{2mk}\rangle_1 = N_{2mk}(|\beta_-\rangle_1 + A_{2mk}|1,odd\rangle_1), \quad (A3)$$
$$|\Psi_{2m+1k}\rangle_1 = N_{2m+1k}(|\beta_+\rangle_1 + A_{2m+1k}|1,even\rangle_1), \quad (A4)$$
$$|\Psi_k\rangle_2 = N_k(r_1 c_{0k}(\beta_1 r_1)|1,-\beta_1 r_1\rangle_1 + t_1 c_{1k}(\beta_1 r_1)|-\beta_1 r_1\rangle_1), \quad (A5)$$

where the normalizations factors $N_{2mk}$ and $N_{2m+1k}$ are determined by analogy with $N_{2m}$ and $N_{2m+1}$ presented above and $N_k = (r_1^2|c_{0k}(\beta_1 r_1)|^2 + t_1^2|c_{1k}(\beta_1 r_1)|^2)^{-1/2}$. The total normalization factors of the state $|\Delta'_{2mk}\rangle_{156}$ and $|\Delta'_{2m+1k}\rangle_{156}$ are given by

$$N^{(t)}_{2mk} = (|a_0|^2 + |a_1|^2|B_{2mk}|^2)^{-1/2}, \quad (A6)$$
$$N^{(t)}_{2m+1k} = (|a_0|^2 + |a_1|^2|B_{2m+1k}|^2)^{-1/2}. \quad (A7)$$

The factors $B_{2mk}$ and $B_{2m+1k}$ determining the entangled properted of two-mode hybrid states can be written as

$$B_{2mk} = t \frac{c_{0k}(\beta_1 r_1)c_{12m}(\beta r)}{c_{02m}(\beta r)} \frac{N^{(0)}_{ev}(\beta t)}{N^{(0)}_{odd}(\beta t)} \frac{N_k}{N_{2mk}}, \quad (A8)$$

$$B_{2m+1k} = t \frac{c_{0k}(\beta_1 r_1)c_{12m+1}(\beta r)}{c_{02m+1}(\beta r)} \frac{N^{(0)}_{odd}(\beta t)}{N^{(0)}_{ev}(\beta t)} \frac{N_k}{N_{2m+1k}}. \quad (A9)$$

The amplitudes $A_{2mk}$ and $A_{2m+1k}$ turns out to coincide with $A_{2mk}$ and $A_{2m+1\ k}$, respectively, in Eqs. (22, 23).

## Appendix B. Analytical expressions for the fidelities between the states (15, 17) and (35-38)

The fidelity between the two types of entangled states, one of which is DV-CV (15, 17) and the other DV-DV (35-38) can be calculated using the definition of the quantity presented above. Here we only provide the final expressions for the fidelities that were used to plot the graphs on Fig. 4. So, the fidelity between the states $|\Delta_0\rangle_{13}$ (Eq. (15)) and $|\Omega^{(2)}_0\rangle_{13}$ (Eq. (35)) is given by

$$F^{(2)}_0 = 4F^2(\beta t)N^{(t)2}_0(|a_0|^2(1 + \beta^4 t^4/2) + |a_1|^2(1 - t^2)(1 + 3\beta^4 t^4/2)). \quad (B1)$$

Similar calculations give the following fidelity

$$F^{(3)}_0 = 4F^2(\beta t)N^{(t)2}_0 \begin{pmatrix} |a_0|^2(1 + \beta^4 t^4/2 + \beta^8 t^8/4!) + \\ |a_1|^2(1 - t^2)(1 + 3\beta^4 t^4/2 + 5\beta^8 t^8/4!) \end{pmatrix}, \quad (B2)$$

between the state $|\Delta_1\rangle_{13}$ (Eq. (17)) and $|\Omega^{(3)}_0\rangle_{13}$ (Eq. (37)). Here, the parameter $N^{(t)}_0$ is the following

$$N_0^{(t)} = \left(|a_0|^2 N_{ev}^{(0)-2}(\beta t) + |a_1|^2 N_0^{-2}\right)^{-1/2}, \tag{B3}$$

with

$$N_0 = \left(r^2 N_{odd}^{(1)-2}(\beta t) + t^2 |c_{10}(\beta r)|^2 N_{odd}^{(0)-2}(\beta t) - 8rt^2 \beta c_{10}(\beta r) exp(-2\beta^2 t^2)\right)^{-1/2}. \tag{B4}$$

Consider again the entangled hybrid state $|\Delta_0\rangle_{13}$ (Eq. (15)) and the state $\left|\Omega_0^{(3)}\right\rangle_{13}$ (Eq. (36)) resulting from truncated superposition (34) with first two terms. Then, the fidelity between the states is calculated

$$F_1^{(2)} = 4F^2(\beta t)N_1^{(t)2}\left(|a_0|^2 t^2(1-t^2)\beta^4 + |a_1|^2 t^2(1+\beta^4(t^2-2r^2)^2/2)\right). \tag{B5}$$

The fidelity between $|\Delta_1\rangle_{13}$ (Eq. (17)) and the state $\left|\Omega_1^{(3)}\right\rangle_{13}$ (Eq. (38)) is given by

$$F_1^{(3)} = 4F^2(\beta t)N_1^{(t)2}\begin{pmatrix}|a_0|^2 t^2(1-t^2)\beta^4(1+\beta^4 t^4/3!) + \\ |a_1|^2 t^2(1+\beta^4(t^2-2r^2)^2/2 + \beta^8 t^4(t^2-4r^2)^2/4!)\end{pmatrix}, \tag{B6}$$

where

$$N_1^{(t)} = \left(|a_0|^2 |c_{01}(\beta r)|^2 N_{odd}^{(0)-2}(\beta t) + |a_1|^2 N_1^{-2}\right)^{-1/2}, \tag{B7}$$

$$N_1 = \left(r^2 |c_{01}(\beta r)|^2 N_{ev}^{(1)-2}(\beta t) + t^2 |c_{11}(\beta r)|^2 N_{ev}^{(0)-2}(\beta t) + \right.$$
$$\left. 8rt^2 \beta c_{01}(\beta r)c_{11}(\beta r)exp(-2\beta^2 t^2)\right)^{-1/2}. \tag{B8}$$

## Appendix C. Interaction of $|\Psi_{2m}\rangle$ with delocalized photon in Fig. 1(a)

In the general case, the calculation of the output state resulting in the interaction of the CV state $|\Psi_{2m}\rangle$ with a delocalized photon (1) is tedious. Consider it on example of the conditional state when even number of photons $2l$ is measured second auxiliary mode. For convenience, we divide the resulting state into two parts. First of them arises from interaction of the CV state $|\Psi_{2m}\rangle_1$ with $|01\rangle_{23}$ on the beam splitter in Fig. 1(a)

$$a_0 F(\beta r)N_{odd}^{(0)-1}(\beta t)R_{2l}(|\beta_-\rangle_1 + C_{2l}|1, odd\rangle_1)|1\rangle_3. \tag{C1}$$

If we consider the interaction of the CV state $|\Psi_{2m}\rangle_1$ with $|10\rangle_{23}$ on the beam splitter in Fig. 1(a), one obtains

$$a_1 F(\beta r)N_{ev}^{(0)-1}(\beta t)R'_{2l}(|\beta_+\rangle_1 + D_{2l}|1, even\rangle_1 + F_{2l}|2, even\rangle_1)|0\rangle_3, \tag{C2}$$

where we introduce the following designations

$$R_{2l} = c_{02l}(\beta r)N_{odd}^{(0)}(\beta) - rA_{2m}c_{12l}(\beta r)N_{odd}^{(1)}(\beta), \tag{C3}$$

$$R'_{2l} = tc_{12l}(\beta r)N_{odd}^{(0)}(\beta) - \sqrt{2}trA_{2m}c_{22l}(\beta r)N_{odd}^{(1)}(\beta), \tag{C4}$$

$$C_{2l} = \frac{tA_{2m}c_{02l}(\beta r)N_{odd}^{(0)}(\beta t)N_{odd}^{(1)}(\beta)}{R_{2l}N_{odd}^{(1)}(\beta t)}, \tag{C5}$$

$$D_{2l} = \frac{N_{ev}^{(0)}(\beta t)\left(rc_{02l}(\beta r)N_{odd}^{(0)}(\beta) + (t^2 - r^2)A_{2m}c_{12l}(\beta r)N_{odd}^{(1)}(\beta)\right)}{R'_{2l}N_{ev}^{(1)}(\beta t)}, \tag{C5}$$

$$F_{2l} = \frac{\sqrt{2}trA_{2m}c_{02l}(\beta r)N_{ev}^{(0)}(\beta t)N_{odd}^{(1)}(\beta)}{R'_{2l}N_{ev}^{(2)}(\beta t)}. \tag{C6}$$

Note that we here introduced a new notation for the superposition of the displaced two photon states (SDTPS) with an amplitude $\beta t$

$$|2, even\rangle = N_{ev}^{(2)}(\beta t)(|2, -\beta t\rangle + |2, \beta t\rangle) = 2F(\beta t)N_{ev}^{(2)}(\beta t)\sum_{m=0}^{\infty}c_{22m}(\beta t)|2m\rangle, \tag{C7}$$

with corresponding normalization factor $N_{ev}^{(2)}$. Using the expressions makes it possible to write expressions for the entangled hybrid states (43, 44). The use of such a technique allows obtaining analytical expressions for the entangled hybrid states (44, 47, 48). A similar technique can be applied if an odd number of photons is detected in auxiliary mode. Amplitudes $C_{2l+1}$, $D_{2l+1}$ and $F_{2l+1}$ can be determined by analogy with $C_{2l}$, $D_{2l}$ and $F_{2l}$

$$R_{2l+1} = c_{02l+1}(\beta r)N_{odd}^{(0)}(\beta) - rA_{2m}c_{12l+1}(\beta r)N_{odd}^{(1)}(\beta), \tag{C8}$$

$$R'_{2l+1} = tc_{12l+1}(\beta r)N_{odd}^{(0)}(\beta) - \sqrt{2}trA_{2m}c_{22l+1}(\beta r)N_{odd}^{(1)}(\beta), \tag{C9}$$

$$C_{2l+1} = \frac{tA_{2m}c_{02l+1}(\beta r)N_{ev}^{(0)}(\beta t)N_{odd}^{(1)}(\beta)}{R_{2l+1}N_{ev}^{(1)}(\beta t)}, \tag{C10}$$

$$D_{2l+1} = \frac{N_{odd}^{(0)}(\beta t)\left(rc_{02l+1}(\beta r)N_{odd}^{(0)}(\beta)+(t^2-r^2)A_{2m}c_{12l+1}(\beta r)N_{odd}^{(1)}(\beta)\right)}{R'_{2l+1}N_{odd}^{(1)}(\beta t)}, \tag{C11}$$

$$F_{2l+1} = \frac{\sqrt{2}trA_{2m}c_{02l+1}(\beta r)N_{odd}^{(0)}(\beta t)N_{odd}^{(1)}(\beta)}{R'_{2l+1}N_{odd}^{(2)}(\beta t)}. \tag{C12}$$

**Acknowledgement**

S.A.P. is supported by Act 211 Government of the Russian Federation, contract № 02.A03.21.0011, while N.B.A. is supported by the National Foundation for Science and Technology Development (NAFOSTED) under project no. 103.01-2017.08.

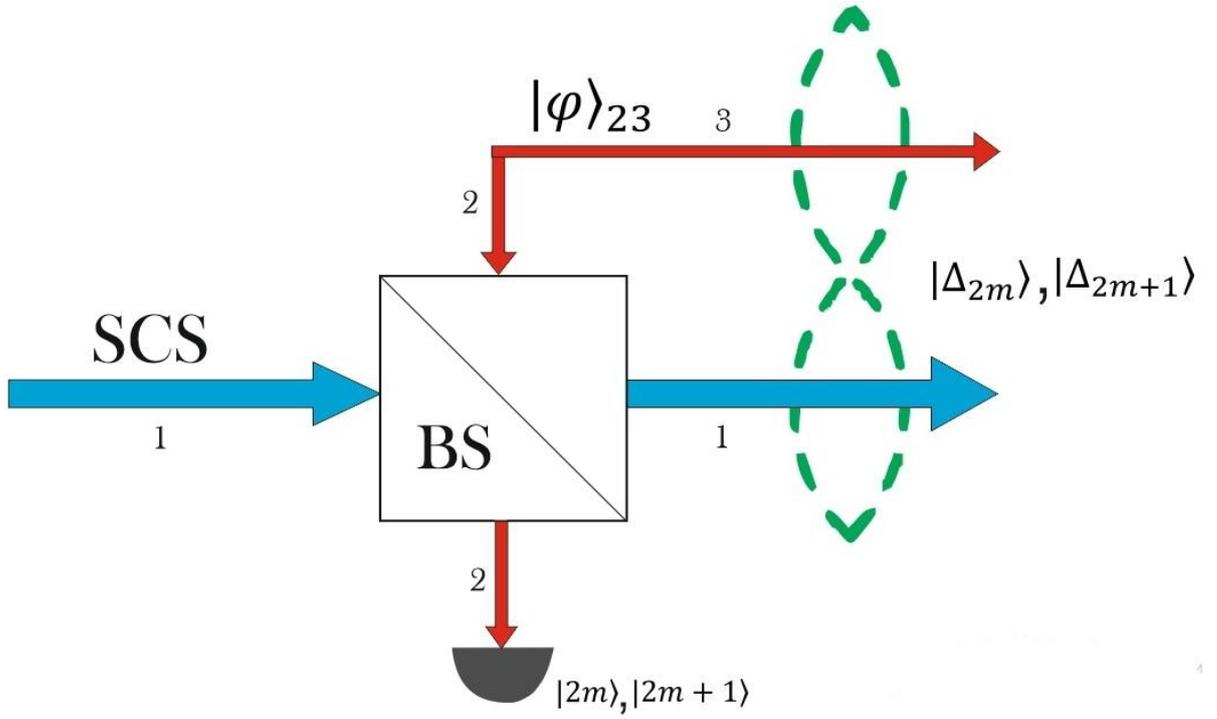

**Fig. 1(a)** Schematic representation of the optical design for generating deterministic entangled hybridity $|\Delta_{2m}\rangle_{13}$ (Eq. (15)) and $|\Delta_{2m+1}\rangle_{13}$ (Eq. (17)) provided that even number of photons $|2m\rangle$ and odd number of photons $|2m+1\rangle$, respectively, are registered in the second mode.

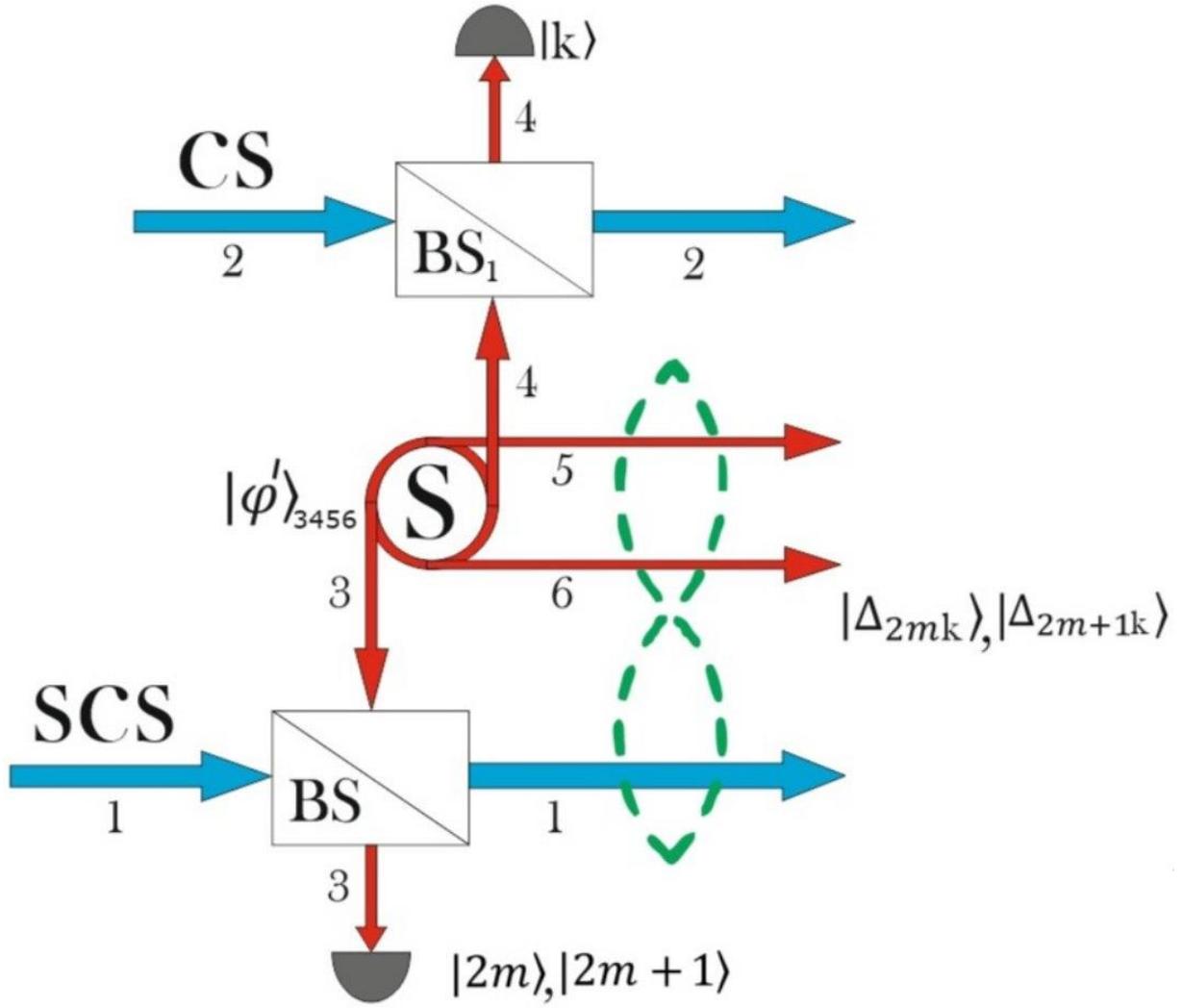

**Fig. 1(b)** Schematic representation of the optical design for generating deterministic entangled hybridity $|\Delta_{2mk}\rangle_{156}$ (Eq. (30)) and $|\Delta_{2m+1k}\rangle_{13}$ (Eq. (31)) provided that even number of photons $|2m\rangle$ and odd number of photons $|2m+1\rangle$, respectively, are registered in the third mode while the detector in fourth mode fixes arbitrary number of photon $|k\rangle$. Two beam splitters $BS$ and $BS_1$ are used to mix SCS and coherent state (CS) with the state (29). $S$ means the source of the states (29).

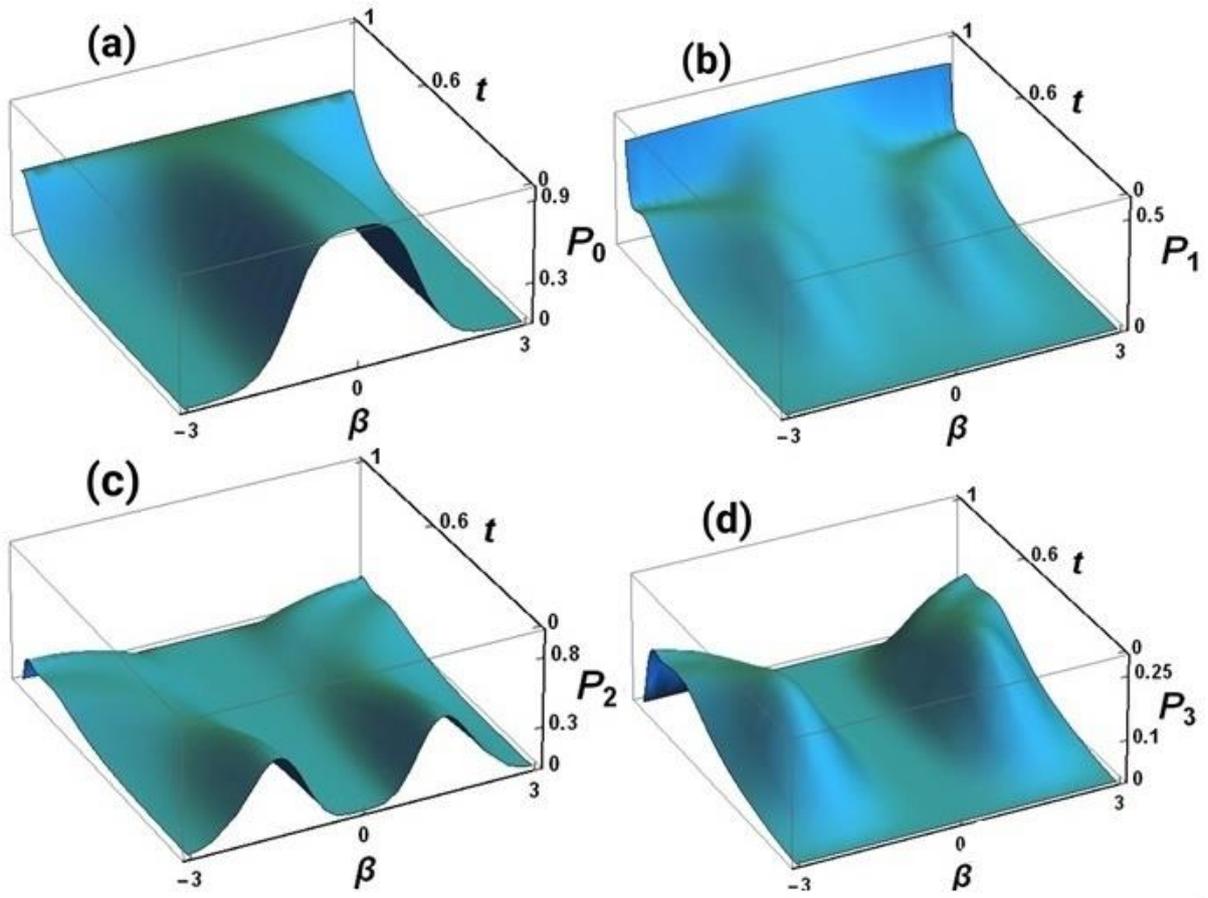

**Fig. 2.** Dependencies of success probabilities: (a) $P_0$, (b) $P_1$, (c) $P_2$ and (d) $P_3$ to generate the hybrid entanglement in Eqs. (15, 17) on $\beta$ and $t$ in the case of $a_0 = a_1 = 1/\sqrt{2}$.

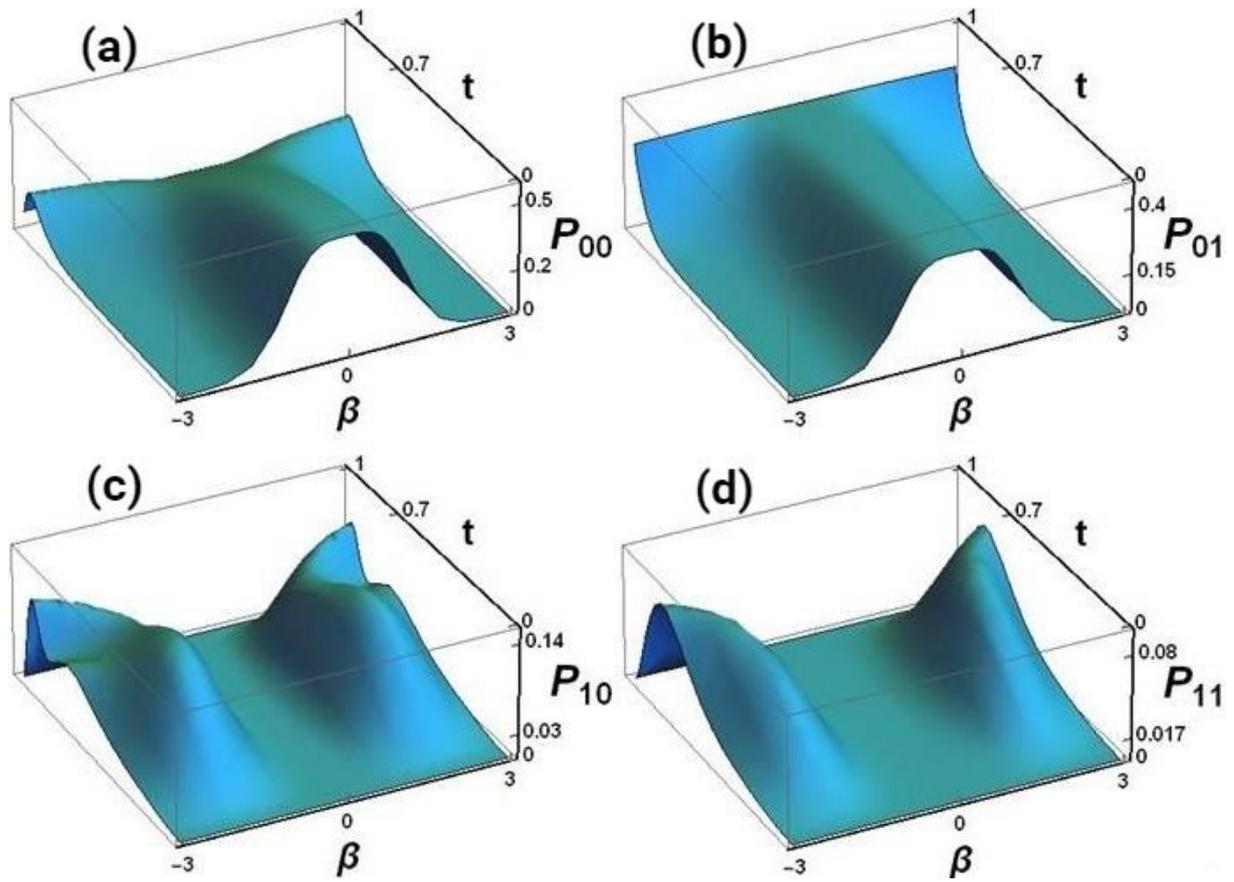

**Fig. 3.** Dependencies of success probabilities: (a) $P_{00}$, (b) $P_{01}$, (c) $P_{10}$ and (d) $P_{11}$ to generate the hybrid entanglement in Eqs. (30, 31) on $\beta$ and $t$ in the case of $a_0 = a_1 = 1/\sqrt{2}$, $\beta_1 = 1$, $t_1 = 0.95$.

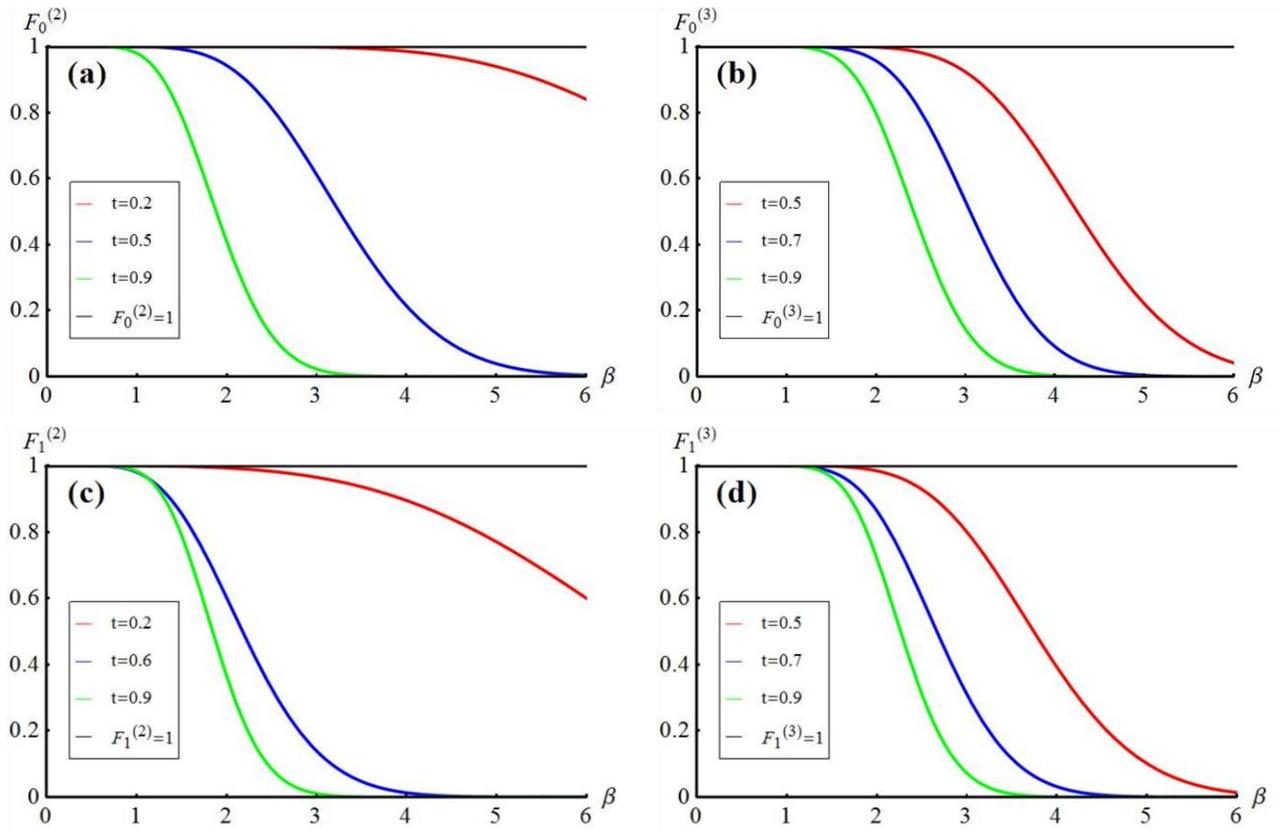

**Fig. 4** Dependencies of the fidelities $F_0^{(2)}$ (a), $F_0^{(3)}$ (b), $F_1^{(2)}$ (c) and $F_1^{(3)}$ (d) between DV-CV (Eqs. 15, 17) and DV-DV (Eqs. (35-38)) entangled states on $\beta$ for different values of the parameter $t$

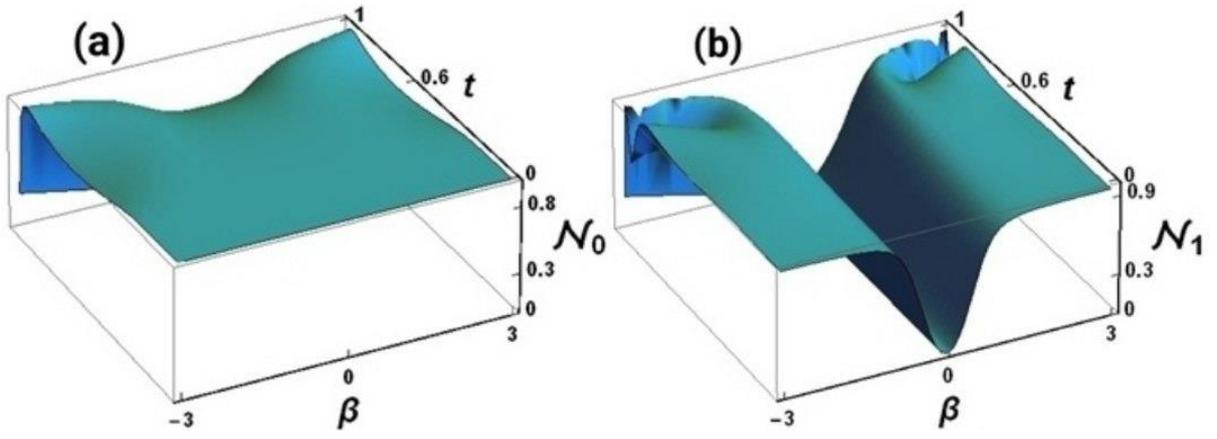

**Fig. 5** Dependencies of measure of entanglement negativity of the conditional states (15, 17): (a) $\mathcal{N}_0$, (b) $\mathcal{N}_1$ on $\beta$ and $t$ in the case of $a_0 = a_1 = 1/\sqrt{2}$.

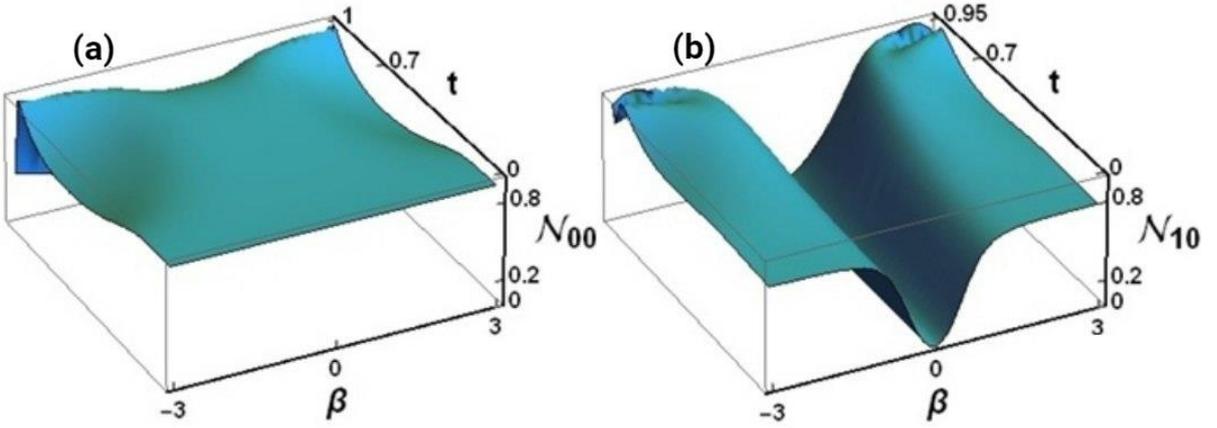

**Fig. 6** Dependencies of measure of entanglement negativity of the conditional states (30, 31): (a) $\mathcal{N}_{00}$, (b) $\mathcal{N}_{10}$ on $\beta$ and $t$ in the case of $a_0 = a_1 = 1/\sqrt{2}$, $\beta_1 = 2$ and $t_1 = 0.95$.